# Application of the Generalized Linear Models in Actuarial Framework

BY MURWAN H. M. A. SIDDIG
*School of Mathematics, Faculty of Engineering Physical Science, The University of Manchester, Oxford Road, Manchester, M13 9PL, U.K.*
murwan_siddig@hotmail.com

*Abstract*—This paper aims to review the methodology behind the generalized linear models which are used in analyzing the actuarial situations instead of the ordinary multiple linear regression. We introduce how to assess the adequacy of the model which includes comparing nested models using the deviance and the scaled deviance. The Akiake information criterion is proposed as a comprehensive tool for selecting the adequate model. We model a simple automobile portfolio using the generalized linear models, and use the best chosen model to predict the number of claims made by the policyholders in the portfolio.

## I. INTRODUCTION

USING the ordinary multiple-linear regression in addressing and analyzing the actuarial situations might be restrictive, as it assumes that the response variable follows the normal distribution only, which is not convenient in practice. The generalized linear models assume a more general class of distributions to the response variable, which makes modeling actuarial situations more feasible.

The generalization we have in the generalized linear models over the ordinary multiple-linear regression is in two matters. The variable of our main interest $y$ which is the one we are trying to explain is allowed to have any distribution that belongs to the exponential family, and not just the Normal distribution. The mean of the variable $y$ doesn't have to be linear on the explanatory variable $x$ $\left(x = (x_1, \ldots, x_p)^T\right)$, as long as it is linear on another scale. To establish notation for the components of the generalized linear models we define the model to have the following system of equations

$$y = \mu + \varepsilon$$
$$g(\mu) = x^T \beta \quad (1)$$

where $y$ is the response variable to some covariate $x = (x_1, \ldots, x_p)^T$. For example, in the actuarial frame-work $y$ could represent the number of claims or the claim sizes produced by a group of policyholders as a response to some risk factors and $\mu = E(Y)$ is the mean response (average number of claims, average claims sizes, . . .). $\varepsilon$ is the random error with mean zero and a constant variance, which is a component that contains the variations in $y$ due to immeasurable random variables. $g(.)$ is the link function, which links the mean response $\mu$ to the linear predictor $\eta = x^T \beta$ and $g(.)$ must be invertible function.

The responses $y_1, \ldots, y_n$ with covariates $x = (x_{i1}, \ldots, x_{ip})^T; i = 1, \ldots, p$ corresponds to different subjects and cases. It is also assumed that the responses are independent, for example the number of claims $y_1$ generated by a certain policyholder is independent of the number of claims $y_2$ generated by another policyholder.

In the Actuarial framework, the parameterization of the model is; for $i = 1, \ldots, n$ we let $y = y_i$ and $x = x_i = (x_{1i}, \ldots, x_{ni})$. The mean $\mu = \mu_i$ and the variance $\mathrm{var}(\mu) = V(\mu_i)\psi_i$ where $\psi_i = \phi/w_i$ and $w_i$ is the weight of the observation $i$, which represents the number of independent and identically distributed observations of which $y_i$ is the arithmetic average for.

## II. THE EXPONENTIAL FAMILY

As we mentioned earlier, in the generalized linear models the distribution of $y$ is assumed to be a member of the exponential family. A distribution is said to be a member of the exponential family if its probability mass/density function can be represented in the form:

$$f(y; \theta, \phi) = \exp\left(\frac{y\theta - b(\theta)}{a(\phi)} + c(y, \phi)\right), \quad (2)$$

where $a$; $b$ and $c$ are given functions. Moreover, $\theta$ and $\phi$ are parameters. In particular, $\theta$ is called the natural parameter, and it is the main parameter of interest. Moreover, $\theta$ is vital in determining the mean of $y$. $\phi$ is called the dispersion parameter and it is vital in determining the variance. $b(\theta)$ is a function that depends only on $\theta$. $a(\phi)$ is a function that depends only on $\phi$. $c(y; \phi)$ is a function that depends on $\phi$ and $y$ but not $\theta$. Moreover, $c(y, \phi)$ does not contribute to the maximum likelihood of $y$ with respect to $\theta$. Further information can be Found in (14).

*Example 1*: Let $y \sim \mathrm{Poisson}(\lambda)$ with probability mass function:

$$f(y; \lambda) = \frac{\lambda^y e^\lambda}{y!}, \quad (3)$$

which can be rewritten as:

$$f(y; \lambda) = \exp\left(\frac{y \log(\lambda) - \lambda}{1} - \log(y!)\right), \quad (4)$$

where $\theta = \log(\lambda)$, $b(\theta) = \lambda$, $a(\phi) = \phi = 1$, and $c(y, \phi) = -\log(y!)$. Additionally, $\lambda = e^\theta \to b(\theta) = e^\theta$ which implies that

$$E(Y) = b'(\theta) = e^\theta = \lambda = \mu$$
$$\text{var}(Y) = b''(\theta)a(\phi) = e^\theta \cdot 1 = \lambda = \mu$$

Furthermore, $g(\mu)$ is the link function (canonical link), to find it we write $\theta$ as a function of $\mu$. Hence, $g(\mu) = \theta = \log(\mu)$ (the log-link function).

*Remark 1:* A very important feature of the generalized linear models is that we can have multiplicative models instead of just having to use additive models like in the ordinary-multiple linear regression. Assume we have that $E(n_{ij}) = \mu \alpha_i \beta_j$. When creating the generalized linear model, we will fix $\alpha_1 = \beta_1 = 1$. Thus, $E(n_{11}) = \mu \alpha_1 \beta_1 = \mu$, which means that $E(n_{11})$ can be interpreted as the expected value of the observation $i = 1$ and $j = 1$. The reason for doing this is to avoid the so called "Dummy-Trap", or " Multi-collinearity", which is a problem that arises when a constant term such as $\mu$ and a factor such as $\alpha$ or $\beta$ are present in the model. In other words, when two or more predictor variables in the model are highly correlated. The problem can be resolved simply by dropping one level of each of the categorical variables, or alternatively dropping the intercept term. Further information can be found in (6).

### III. GOODNESS-OF-FIT
*A. General*

To find a model which fits the data adequately, where the observed value doesn't differ so much from the predicted value we look at the residuals. The smaller the residuals, the better the model. The model with the smallest residuals is the so-called"Full model", where every observation $i$ has its own parameter and the model is simply repeating the data, rather than predicting it. The model with the largest residuals is the so-called"Null-model" where the model ascribes all the variations in $y$ to a constant term without considering the collateral data. The null model is too crude and the full model has too many parameters for practical use. The optimal model is somewhere between those two extremes. To check the residuals, we look at the deviance $D$ or the scaled deviance $D^*$. In this paper we are going to use the scaled deviance. If $D > \mathcal{X}^2_{(\alpha; n-p)}$ we reject the null hypothesis $H_0$: the residuals deviance is not significantly large and the model is good as far as the residuals is concerned, where $\mathcal{X}^2_{(\alpha; n-p)}$ is the Chi-square test value at significance level $\alpha$ and $p$ parameters. If not, otherwise. We also look at the significance of the estimated parameters by doing another hypothesis test.

*B. Comparing Nested Models*

After fitting a generalized linear model by checking the residuals and checking the significance of the parameters we can add some refinements to the model. For example, we can add an interaction term if we are using an additive model, or we can eliminate some parameters. This is done by using the so-called" Change in scaled deviance" $D_0^* - D_1^*$, where $D_0^*$ is the scaled deviance of the new model, and $D_1^*$ is the scaled deviance of the old model. If $D_0^* - D_1^* > \mathcal{X}^2_{(\alpha; q)}$ where $q$ is the difference between the number of parameters of the new model and the number of parameters of the old model, then the new model is rejected. If not, otherwise.

*C. Comprehensive Criterion for Model selection*

It is very likely that we end up with several nested models where the inclusion or the exclusion of one parameter leads to another model. The inclusion of new parameters usually reduces the residuals. Nonetheless, the complexity which arises is assessing whether the reduction in the residuals justifies the inclusion of the new parameter or not. This could be resolved by using the Akiake Information criterion (AIC). The best model is the one which has the smallest AIC.

### IV. EXAMPLE

In this section we are going to propose a generalized linear model to analyze a simple automobile portfolio.

Assume we observed drivers in some fictional motor insurance portfolio, and recorded the number of claims produced by those policyholders over the period of 7 years. Additionally, we classified the drivers according to the risk factors sex, region, job class and type of car.

Table I
OBSERVED NUMBER OF CLAIMS AND EXPOSURE IN THE PORTFOLIO

| Sex | Region | Type of Car 1 | | | 2 | | | 3 | | |
|---|---|---|---|---|---|---|---|---|---|---|
| | | Job 1 | 2 | 3 | 1 | 2 | 3 | 1 | 2 | 3 |
| 1 | 1 | 14 | 52 | 48 | 104 | 48 | 79 | 80 | 69 | 68 |
| | 2 | 51 | 38 | 78 | 98 | 82 | 113 | 116 | 95 | 118 |
| | 3 | 69 | 87 | 43 | 88 | 88 | 129 | 148 | 112 | 125 |
| 2 | 1 | 52 | 75 | 44 | 89 | 33 | 93 | 66 | 92 | 95 |
| | 2 | 86 | 24 | 71 | 89 | 53 | 77 | 75 | 170 | 127 |
| | 3 | 91 | 109 | 69 | 104 | 113 | 179 | 117 | 149 | 133 |

Our main aim is to establish a tariff to the portfolio. Hence, we want to predict the number of claims each policyholder is expected to make. The technique in doing so is to relate the annual number of claims frequency to the risk factors by estimating a parameter for each risk factor. To establish notation for the proposed generalized linear model we let $n_{ijkl}$ be the observed number of claims made by the policyholders in the cell *ijkl* where, *i, j, k* and *l* represents the different levels of the risk factors sex, region, type of car, and job class, consecutively. Additionally, $i = 1,2$. $j = 1,2,3$. $k = 1,2,3$. $l = 1,2,3$.

In general, $n_{ijkl} \sim \text{Binomial}(w_{ijkl}, \pi_{ijkl})$, where $\pi_{ijkl} = N\pi_{i...}\pi_{.j..}\pi_{..k.}\pi_{...l}$, is the probability that the policyholders in the cell *ijkl* are going to make a claim, $N$ is the total number of policyholders in the portfolio and $w_{ijkl}$ is the number of policyholders exposed to risk in the cell *ijkl*. Nonetheless, because $w_{ijkl}$ is large and $\pi_{ijkl}$ is small, we are going to approximate this Binomial distribution by a Poisson distribution, such that $n_{ijkl} \sim \text{Poisson}(\mu_{ijkl})$ with a Log-link function $g(\mu_{ijkl}) = \log(\mu_{ijkl})$ (see Example 1), as shown in Yuan, J (2014). Hence, the generalized linear model we are going to use is:

$$n_{ijkl} = \mu_{ijkl} + \varepsilon_{ijkl} \sim \text{Poisson}(\mu_{ijkl}),$$
$$g(\mu_{ijkl}) = \log(\mu_{ijkl}). \quad (5)$$

But $\mu_{ijkl} = w_{ijkl}\pi_{ijkl}$. This means that $g(\mu_{ijkl}) = \log(w_{ijkl}\pi_{ijkl}) = \log(w_{ijkl}) + \log(\pi_{ijkl})$.

The first model we are going to start with is Fit1:
$$g(\mu_{ijkl}) = \mu + \alpha_i + \beta_j + \gamma_k + \delta_l \quad (6)$$

Where $\mu$ is the intercept term, $\alpha_i, \beta_j, \gamma_k,$ and $\delta_l$ are the Parameters representing the risk factors sex, region, type of car, and job, consecutively.
Fitting the model using $R$ we get the results shown in Table II.

Table II
ESTIMATES OF THE PARAMETERS OF THE MODEL FIT1

|  | Estimate | Std. Error | z value | pr(> \|z\|) |
|---|---|---|---|---|
| (Intercept) | -3.09959 | 0.12294 | -25.211 | $< 2e^{-16}$*** |
| sex2 | 0.10303 | 0.07634 | 1.350 | 0.17710100 |
| region2 | 0.23468 | 0.09924 | 2.365 | 0.0180450* |
| region3 | 0.46434 | 0.09655 | 4.809 | $1.51e^{-06}$*** |
| type2 | 0.39463 | 0.10168 | 3.881 | 0.000104*** |
| type3 | 0.58443 | 0.09709 | 6.019 | $1.75e^{-09}$*** |
| job2 | -0.03617 | 0.09695 | -0.373 | 0.70906000 |
| job3 | 0.06072 | 0.09261 | 0.656 | 0.51208100 |

Additionally, we get that the null deviance is 104.73 on 53 degrees of freedom. The residuals deviance is 41.93 on 46 degrees of freedom, and the AIC is 288.24.
To check the residuals, we do a hypothesis test $H_0$: The residuals deviance is not significantly large and the model is good as far as the residuals are concerned. $H_0$: Otherwise. Since $D^*$=41.93 on 46 degrees of freedom, then $X^2_{41.93; 46} = 0.6433506 > 0.05$. Hence, we have no evidence to reject $H_0$ at 5% significance level and the model is good as far as the residuals are concerned.
Furthermore, to test the significance of the parameters, we look at their pr(> |z|) values in Table II. We can see that the parameters sex2, job2, and job3 are insignificance at 5% significance level, because their pr(|z|) values are greater than 0:05. Accordingly, they can be dropped from the model. Nevertheless, it might be risky to drop them both at once. Therefore, in the newly proposed model Fit2 we are going to drop only the risk factor sex. Hence, Fit2 is:

$$g(\mu_{jkl}) = \mu + \beta_j + \gamma_k + \delta_l \quad (7)$$

Fitting the model using $R$ we get the results shown in Table III. Additionally, we get that the null deviance is 104.732 on 53 degrees of freedom, the Residual deviance is 43.755 on 47 degrees of freedom, and the AIC is 288.06.

Table III
ESTIMATES OF THE PARAMETERS OF THE MODEL FIT2

|  | Estimate | Std. Error | z value | pr(> \|z\|) |
|---|---|---|---|---|
| (Intercept) | -3.05043 | 0.11716 | -26.036 | $< 2e^{-6}$*** |
| region2 | 0.23673 | 0.09922 | 2.386 | 0.01700000* |
| region3 | 0.46460 | 0.09655 | 4.812 | $1.49e^{-06}$ *** |
| type2 | 0.39849 | 0.10163 | 3.921 | $8.82e^{-05}$ *** |
| type3 | 0.58431 | 0.09709 | 6.018 | $1.76e^{-09}$*** |
| job2 | -0.03253 | 0.09692 | -0.336 | 0.73700000 |
| job3 | 0.06478 | 0.09255 | 0.700 | 0.48400000 |

To check the residuals we do the same hypothesis test we did in testing the model Fit1. $H_0$: The residuals deviance is not significantly large and the model is good as far as the residuals are concerned. $H_1$: Otherwise. Since $D^*$= 43.755 on 47 degrees of freedom, then $X^2_{(43.755; 47)} = 0.6077625 > 0.05$. Hence, we have no evidence to reject $H_0$ at 5% significance level and the model is good as far as the residuals are concerned.
To test the significance of the parameters, we look at their pr(> |z|) values in Table III. We can see that the parameters for the risk factor job remains insignificance at 5% significance level, because their pr(|z|) values are greater than 0.05. Hence, they can be dropped from the newly proposed model Fit3:

$$g(\mu_{jk}) = \mu + \beta_j + \gamma_k \quad (8)$$

Fitting the model using $R$ we get the results shown in Table IV.

Table IV
ESTIMATES OF THE PARAMETERS OF THE MODEL FIT3

|  | Estimate | Std. Error | z value | pr(> \|z\|) |
|---|---|---|---|---|
| (Intercept) | -3.03132 | 0.10151 | -29.862 | $< 2e^{-16}$ *** |
| region2 | 0.23141 | 0.09905 | 2.336 | 0.01947500* |
| region3 | 0.46046 | 0.09647 | 4.773 | $1.82e^{-06}$*** |
| type2 | 0.39419 | 0.10149 | 3.884 | 0.000103*** |
| type3 | 0.58331 | 0.09707 | 6.009 | $1.86e^{-09}$*** |

Additionally, we get that the null deviance is 104.73 on 53 degrees of freedom, the Residual deviance is 44.94 on 49 degrees of freedom, and the AIC is 285.25. To check the residuals we do the same hypothesis test we did in testing Fit1 and Fit2. $H_0$: The residuals deviance is not significantly large and the model is good as far as the residuals are concerned. $H_1$: Otherwise. Since $D^* = 44.94$ on 49 degrees of freedom, then $\chi^2_{44.94,49} = 0.6383855 > 0.05$. Hence, we have no evidence to reject $H_0$ at 5% significance level and the model is good as far as the residuals are concerned. Moreover, looking at the $pr(>|z|)$ values in Table IV we can see that all the parameters in the model Fit3 are significance at 5% significance level.

The same results could have been deduced using the AIC values. Since $285.25 < 288.06 < 288.24$ then the best model is Fit3.

It is also very instructive to try and see if there is any interaction between the significant risk factors, namely region and type of car. This can be done by adding an interaction term $\eta_{jk}$ to the model, such that the newly proposed model Fit4 is:

$$g(\mu_{jk}) = \mu + \beta_j + \gamma_k + \eta_{jk} \qquad (9)$$

Fitting the model using R we get the results shown in Table V.

Table V
ESTIMATES OF THE PARAMETERS OF THE MODEL FIT

| | Estimate | Std. Error | z value | pr(>\|z\|) |
|---|---|---|---|---|
| (Intercept) | -2.98873 | 0.15250 | -19.598 | $< 2e^{-16}$*** |
| region2 | 0.14988 | 0.20612 | 0.727 | 0.4671 |
| region3 | 0.42165 | 0.19273 | 2.188 | 0.0287* |
| type2 | 0.43376 | 0.19846 | 2.186 | 0.0288* |
| type3 | 0.45195 | 0.19273 | 2.345 | 0.0190* |
| region2:type2 | -0.08084 | 0.26639 | -0.303 | 0.76150 |
| region3:type2 | -0.02230 | 0.25374 | -0.088 | 0.93000 |
| region2:type3 | 0.25559 | 0.25619 | 0.998 | 0.31850 |
| region3:type3 | 0.10860 | 0.24487 | 0.443 | 0.65740 |

Additionally, we get that the null deviance is 104.73 on 53 degrees of freedom, the Residual deviance is 42.412 on 45 degrees of freedom, and the AIC is 290.72.

Looking at the newly added term $pr(>|z|)$ values shown in Table V, we can see that the interaction term is insignificant at 5% significance level. Hence, it should not be included in the model, and the best model is Fit3.

The same results could be deduced using the AIC values. Since $285.25 < 288.06 < 288.24 < 290.72$ then the best model is Fit3.

For prediction we will use the best estimated model Fit3. Combining Fit3 and equation 5 we have that $E(n_{jk}) = \hat{\mu}_{jk}$ where $\log(\mu_{jk}) = \mu + \beta_j + \gamma_k$. Consequently, $\mu_{jk} = e^{\mu+\beta_j+\gamma_k}$ and $\hat{\mu}_{jk} = e^{\widehat{\mu}+\widehat{\beta_j}+\widehat{\gamma_k}}$. We let $\beta_1 = \gamma_1 = 0$ to avoid the Dummy-Trap as shown in Remark 1. Thus, we will have the following:

$$n_{11} = e^{\mu+\beta_1+\gamma_1} = e^{-3.03132} = 0.048252. \qquad (10)$$

To know how many years it will take the policyholder of the corresponding cell in Equation 10 to make one claim, we have that $0.048252 * x = 1$, then $x = 1/0.048252 = 20.725$. Similarly

$$n_{12} = e^{\mu+\beta_1+\gamma_2} = e^{-3.03132+0.39419} = 0.0715663. \qquad (11)$$

Accordingly, the number of years it will take the policyholder of the corresponding cell in Equation 11 to make one claim is $0.0715663 * x = 1$, then $x = 1/0.0715663 = 13.973057$. Similarly, we can do the same for $n_{13}, n_{21}, n_{22}, n_{23}, n_{31}, n_{32}$, and $n_{33}$.

V. CONCLUSION

A very important matter that actuaries should treat carefully is the rating system they set to calculate the premiums to be taken from the policyholders. In non-life insurance in particular, policyholders may leave when they think they are overcharged. In contrast, a wrong rating system may attract bad risk. The ultimate assessment to see how accurate the rating system is, how precise it reflects the observed losses. Classifying the observed losses according to the appropriate risk factors is very substantial in determining how accurate the rating system is, in the sense that, the risk factors tell us exactly which level of which risk factor causes the biggest loss (to be charged the highest premium), and which causes the smallest loss (to be charged the lowest premium). Apart from the general risk factors (region of residence, age of the policyholder, type of usage ...), some insurance companies tend to classify the observed losses according to the so called"Bonus-Malus system". The Bonus-Malus system is a system in which a claim-free year (a year without reporting any claims) leads to a discount in the premium (bonus). On the other hand, the premium increases as a consequence of a year with bad claim record (malus). A sketched table of the Bonus-Malus system is given in the Appendix. The first thing which should be pointed in the Bonus- Malus system is that step 1 (120%) is the only malus class in the system. Furthermore, a policyholder without any known history (new customer to the company) starts from step 2 (100%). To elaborate further, in general, insurance companies have some sort of basic premium (premium factor), where the premium each policyholder is going to pay adjusts according to. For instance, assume that the basic premium is $500, a policyholder without any known history will be charged 100% of this basic premium i.e. $500. Moreover, a policyholder in the only malus step (step 1) will be charge 120% of this premium i.e. $600. Nonetheless, in practice there are certain behaviors policyholders have towards the Bonus-Malus system, more precisely in their desire to get a bonus (discount). For example, the longer a policyholder has remained in the highest bonus level namely step 14, the less hungry for a bonus he/she is, and they are more willing to report small claims. This is called the Bonus-Guarantee. The convenient explanation to this is that, if a claim will not lead to more premiums having to be paid, small claims will be filed as well, leading to more and somewhat smaller claims than other classes. The reason why we are introducing this concept of bonus guarantee is that, when looking at

the pattern of the relationship between the losses caused by the policyholders and their step in the bonus-malus system, what we would expect to see is that, the higher step the policyholder is in, the less losses he/she is causing. This is because in order for you to get promoted to the next step you must have not had any claims filed in the previous year. However, the pattern is normally consistent from the bonus steps 1-13 but not 14, this is due to the bonus guarantee, and it should be treated carefully when modeling the portfolio.

Actuaries should also be aware of the so-called "The danger of the one-dimensional analysis". The actuary should not be tempted to stop the analysis in finding the averages of responses caused by each risk factor in the portfolio. For instance, saying that the average number of claims expected to be made by a policyholder who is living in the rural areas, driving an average of 1-7500 mileage per year is the sum of the average numbers of claims caused by a policyholder who is living in the rural areas, driving an average of 1-7500 mileage per year. The reason why doing this might be very risky is, these risk factors are very likely to be correlated and hence the actuary will fall into the danger of the one-dimensional analysis". To illustrate further, typically in the rural areas there is not a lot of traffic, so there are few claims. In contrast, in big cities there is traffic, so there are a lot of claims. Furthermore, drivers with low mileage tend to be inexperienced, which means more claims. Whereas, drivers with high mileage tend to drive in high ways, which means experienced drivers and hence less claims. However, the mileage driven by the policyholder and the region he/she is living in are very likely to be correlated (traffic means driving in crowded places and hence driving for small distance i.e. low mileage, and vice versa). Thus, drivers who are living in the big cities might be charged more, both for living in the big city and for driving for small number of mileages, even though one factor is correlated with the other one (one factor is causing the other one to happen). Another way to look at this is, if we included the bonus-malus system in our analysis, one might see that young drivers might be charged more for both being inexperienced and for the fact that they cannot be in high classes yet.

Just like any other extended actuarial context, whether it was additional observations or additional risk factors, the analysis of the Bonus-Malus system in the Generalized linear models is done by defining the Bonus-Malus class the policyholder in, as a new risk factor.

## ACKNOWLEDGMENT

I thank Dr. Kees Van Schaik for the helpful comments, and taking the time and effort to supervise the full version of this project.

## APPENDIX

Table VI
TRANSITION RULES AND PREMIUM PERCENTAGES FOR A BONUS-MALUS SYSTEM

| step | 1 | 2 | 3 | 4 | 5 | 6 | 7 | 8 | 9 | 10 | 11 | 12 | 13 | 14 |
|---|---|---|---|---|---|---|---|---|---|---|---|---|---|---|
| percentages | 120 | 100 | 90 | 80 | 70 | 60 | 55 | 50 | 45 | 40 | 37.5 | 35 | 32.5 | 30 |
| 0 claims → | 2 | 3 | 4 | 5 | 6 | 7 | 8 | 9 | 10 | 11 | 12 | 13 | 14 | 14 |
| 1 claim → | 1 | 1 | 1 | 1 | 2 | 3 | 4 | 5 | 6 | 7 | 7 | 8 | 8 | 9 |
| 2 claims → | 1 | 1 | 1 | 1 | 1 | 1 | 1 | 1 | 2 | 3 | 3 | 4 | 4 | 5 |
| 3 claims → | 1 | 1 | 1 | 1 | 1 | 1 | 1 | 1 | 1 | 1 | 1 | 1 | 1 | 1 |

## REFERENCES

[1] Andrews, D. W. (1991). *Heteroskedasticity and autocorrelation consistent covariance matrix estimation*. Econometrica: Journal of the Econometric Society, 817-858.
[2] Antonio, K., & Beirlant, J. (2007). *Actuarial statistics with generalized linear mixed models*. Insurance: Mathematics and Economics, 40(1), 58-76.
[3] Boshnakov, G. N. (2007). *Lecture notes for Time Series Analysis and Financial Forecasting*, MSc notes. The University of Manchester.
[4] De Jong, P., & Heller, G. Z. (2008). *Generalized linear models for insurance data* (Vol. 136). Cambridge: Cambridge University Press.
[5] Denuit, M., & Dhaene, J. (2001). *Bonus-Malus scales using exponential loss functions*. Blätter der DGVFM, 25(1), 13-27.
[6] Farrar, D. E., and Glauber, R. R. (1967). *Multicollinearity in regression analysis: the problem revisited*. The Review of Economic and Statistics, 92-107.
[7] Fox, J. (2015). *Applied regression analysis and generalized linear models*. 3rd edn. Sage Publications.
[8] Kaas, R., Goovaerts, M., Dhaene, J., and Denuit, M. (2008). *Modern actuarial risk theory: using R* (Vol. 128). Springer Science and Business Media.
[9] Lancaster, H. O. (1969). *Chi-Square Distribution. John Wiley and Sons, Inc.*.
[10] Lange, K. (2010). *Applied probability*. Springer Science and Business Media.
[11] Loeffen, R. (2014). *Lecture notes for General Insurance, The collective risk model*, 97–99. The University of Manchester.
[12] Nelder J.A. & Wedderburn, R.W.M. (1972). *Generalized Linear Models*, Journal of the Royal Statistical Society, A, 135, 370–384.
[13] Verrall R. (1996). *Claims reserving and generalized additive models, Insurance*: Mathematics & Economics 19, 31–43.
[14] Yuan, J. (2014). *Lecture notes for Generalised Linear Models, Goodness-of-fit*. The University of Manchester.